# Gravitational Microlensing Results from MACHO

W. Sutherland [a], C. Alcock [b], R. Allsman[b], T. Axelrod[b], D. Bennett[b], S. Chan [c], K. Cook[b], K. Freeman[c], K. Griest [d], S. Marshall[d], S. Perlmutter[d], B. Peterson[c], M. Pratt[d], P. Quinn[c], A. Rodgers[c], C. Stubbs[d]

[a] Dept. of Physics, 1 Keble Road, Oxford OX1 3RH, U.K.

[b] Lawrence Livermore National Laboratory, Livermore, CA 94550.

[c] Mt. Stromlo & Siding Spring Observatories, Canberra, ACT 2611, Australia.

[d] Center for Particle Astrophysics, Univ. of California, Berkeley, CA 94720.

31 Aug 94 : Talk given at Neutrino-94, to appear in Nucl. Phys. B Conference Supp.

We provide a status report on our search for dark matter in our Galaxy in the form of massive compact halo objects (MACHOs), using gravitational microlensing of background stars. This search uses a very large CCD camera on the dedicated 1.27m telescope at Mt. Stromlo, Australia, and has been taking data for 2 years. At present, we have analysed data for 8 million stars in the Large Magellanic Cloud over 1 year, resulting in one strong candidate event and two lower-amplitude candidates. We have also analysed 5 million stars in the Galactic Bulge for 0.5 years, yielding approximately 27 microlensing events.

## 1. INTRODUCTION

Evidence for large quantities of dark matter in the Universe has now been accumulating for many years [1], and is reviewed by Gould (this volume).

This dark matter cannot be in the form of normal stars or gas, but a wide range of candidates have been proposed. These fall into two main classes: the particle-physics candidates such as massive neutrinos, axions or other weakly interacting massive particles (WIMPs), and the astrophysical candidates such as substellar objects below the hydrogen burning threshold $\approx 0.1 M_\odot$ ('brown dwarfs'), or stellar remnants such as white dwarfs, neutron stars or black holes; these are generically known as massive compact halo objects or MACHOs.

Such objects would be much too faint to have been detected in current sky surveys (though planned infrared facilities such as the SIRTF satellite and 8-metre telescopes may reach the required sensitivity[2]); but Paczynski[3] suggested that MACHOs could be detected by their gravitational 'microlensing' of background stars.

### 1.1. Gravitational Microlensing

If a compact object passes very close to the line of sight to a background star, the light will be deflected to produce two images of the star. In the case of perfect alignment, the star will appear as an 'Einstein ring' with a radius (in the lens plane) of

$$r_E = \sqrt{\frac{4GMLx(1-x)}{c^2}}, \tag{1}$$

where $M$ is the lens mass, $L$ is the observer-source distance and $x$ is the ratio of the observer-lens and observer-source distances. In a typical situation of imperfect alignment, the image appears as two arcs. On cosmological scales, many cases of quasars multiply imaged by foreground galaxies are known ; however, for stellar lensing the angular separation of the lensed images is much smaller. For a source distance of $50\,\mathrm{kpc} \approx 10^{10}\,\mathrm{AU}$ and a lens distance of $10\,\mathrm{kpc}$, the Einstein radius is $r_E \approx 8\sqrt{M/M_\odot}\,\mathrm{AU}$ [1] ; this yields an angular separation of $\sim 0.001$ arcsecond which is well below even the Space Telescope resolution, hence the term 'microlensing'.

[1] We use standard astronomical notation; $M_\odot$ is the mass of the Sun, $1\,\mathrm{AU}$ is the Earth-Sun distance, 1 parsec $\approx 206000\,\mathrm{AU}$.

However, in the point source approximation, the lensing produces a net amplification of the source by a factor

$$A = \frac{u^2 + 2}{u\sqrt{u^2+4}} \quad (2)$$

where $u = b/r_E$ and $b$ is the distance of the lens from the direct line of sight. (The above scale of 8 AU is much larger than the radius of a star, hence the point source approximation is accurate for lens masses $\gtrsim 10^{-5} M_\odot$). The amplification is approximately $u^{-1}$ for $u \lesssim 0.5$, and $1 + 2u^{-4}$ for $u \gg 1$, hence the amplification can be very large but falls rapidly for $u \gtrsim 1$. Since objects in the Galaxy are in relative motion, this amplification will be time-dependent; for a typical lens transverse velocity of $200\,\mathrm{km\,s^{-1}}$, the duration is

$$\hat{t} \equiv \frac{2r_E}{v_\perp} \approx 140\sqrt{M/M_\odot}\ \text{days}. \quad (3)$$

This is a convenient timescale for astronomical observations, and thus by sampling on a range of timescales the microlensing searches may be sensitive to a wide mass range from small planets of $\sim 10^{-6} M_\odot$ to black holes of $\sim 100 M_\odot$, covering most of the plausible candidates.

### 1.2. Optical Depth

The 'optical depth' $\tau$ for gravitational microlensing is defined as the probability that a given star is lensed with $u < 1$ or $A > 1.34$ at any given time, and is

$$\tau = \pi \int_0^L \frac{\rho(l)}{M} r_E^2(l)\, dl \quad (4)$$

where $l$ is the distance along the line-of-sight and $\rho$ is the dark matter density. Since $r_E \propto \sqrt{M}$, while for a given $\rho$ the number density of lenses $\propto M^{-1}$, the optical depth is independent of the individual MACHO masses. Using the virial theorem, it is found that $\tau \sim (v/c)^2$, where $v$ is the rotation speed of the Galaxy. More detailed calculations[4] give an optical depth for lensing by halo dark matter of stars in the Large Magellanic Cloud of

$$\tau_{\rm LMC} \approx 5 \times 10^{-7} \quad (5)$$

This very low value is the main difficulty of the experiment; only one star in two million will be amplified by $A > 1.34$ at any given time, while the fraction of variable stars is much higher, $\sim 3 \times 10^{-3}$.

### 1.3. Microlensing Signatures

Fortunately, microlensing has many strong signatures which can discriminate it from stellar variability :

1) Since the optical depth is so low, only one event should be seen in any given star.
2) The deflection of light is wavelength-independent, hence the star should not change colour during the amplification.
3) The accelerations of galactic objects are negligible on timescales of these events, hence the events should be symmetrical in time, and have a shape derived from (2) with $u(t) = \sqrt{u_{min}^2 + (v_\perp(t - t_{max})/r_E)^2}$.

Examples of such light-curves are shown in Figure 1.

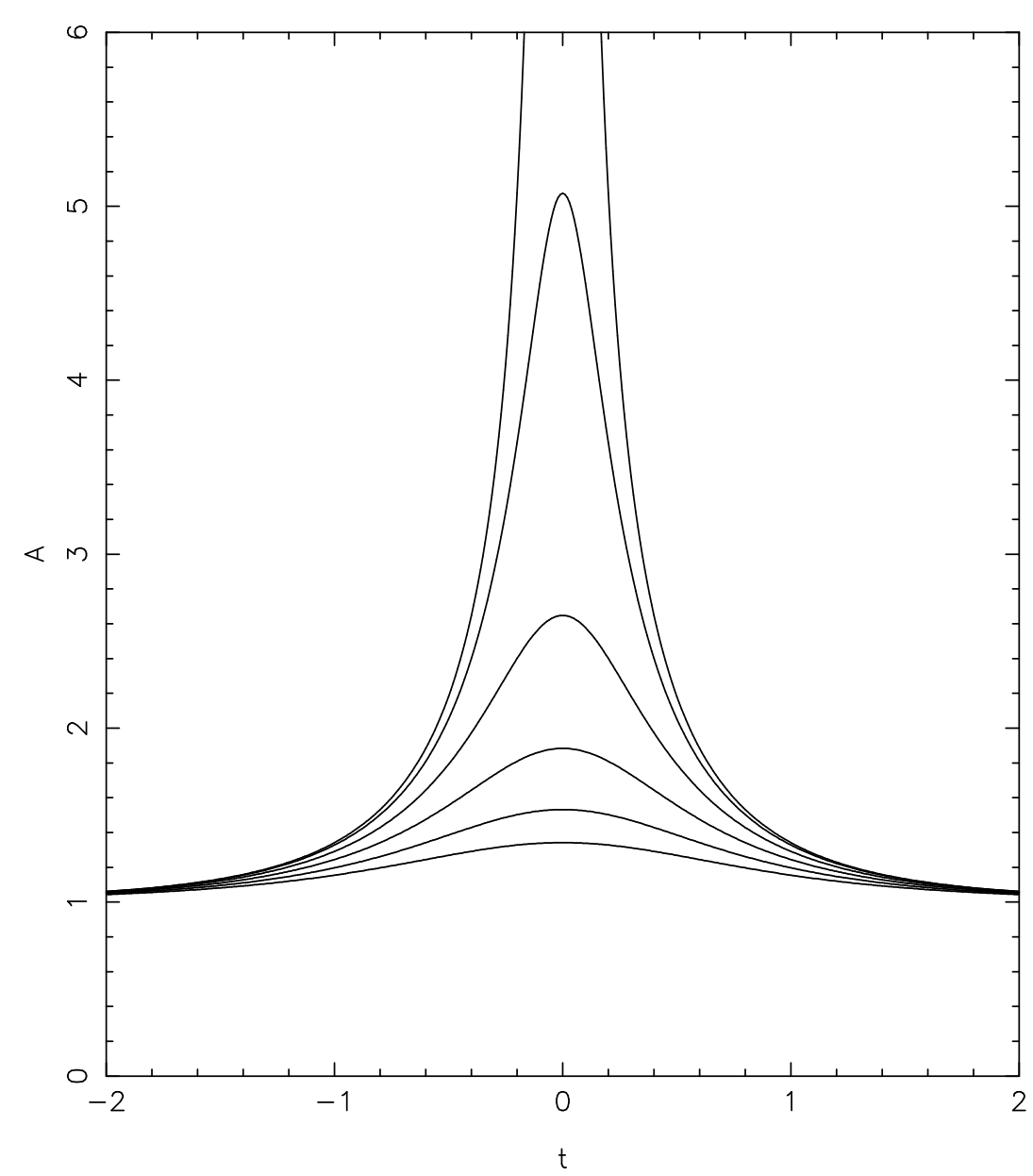

Figure 1. Theoretical microlensing light curves for impact parameters $u_{min} = 0, 0.2, \cdots 1$. Amplification is in linear units, timescale is in units of $r_E/v_\perp$.

All these characteristics are distinct from known types of intrinsic variable stars; most variable stars are periodic or semi-regular, and do not remain constant for long durations. They usually change temperature and hence colour as they vary, and they usually have asymmetrical lightcurves with a rapid rise and slower fall.

In addition to these individual criteria, if many candidate microlensing events are detected, there are further statistical tests that can be applied:

4) The events should occur equally in stars of different colours and luminosities.
5) The distribution of impact parameter $u_{min}$ should be uniform from 0 to the experimental cutoff $u(A_{min})$.
6) The event timescales and peak amplifications should be uncorrelated.

(In practice these distributions will be modulated by the detection efficiencies, which can be computed from simulations).

It is worth clarifying two common misconceptions: first, this experiment does *not* provide an estimate of $\Omega$, since the amount of dark matter within 50 kpc of spiral galaxies is approximately known from rotation curve data. This represents $\Omega \sim 0.05$: if $\Omega = 1$ either the halos must extend far beyond 50 kpc or there must be intergalactic dark matter. Secondly, the lenses are not required to be dark, merely significantly fainter than the source stars. However, since the LMC is located $30^o$ from the galactic plane, the optical depth from known stars is only $\sim 10^{-8}$.

## 2. OBSERVATIONS

Due to the low optical depth, a very large number of stars must be monitored over a long period to obtain significant results. The optimal targets for this search are the Large and Small Magellanic Clouds, the largest of the Milky Way's satellite galaxies, since they have a high surface density of stars and are distant enough at 50 and 60 kpc to provide a good path length through the dark halo. This requires a Southern hemisphere observatory.

The MACHO collaboration has full-time use of the 1.27-m telescope at Mt.Stromlo Observatory near Canberra, Australia. The telescope has been re-commissioned for the project; an optical corrector gives a field of view of $0.7 \times 0.7$ degrees at prime focus, and a dichroic beamsplitter is used to take simultaneous images in red and blue passbands. The two foci are equipped with very large CCD cameras[5], each containing 4 Loral CCD chips of $2048 \times 2048$ pixels. The images are read out through a 16-channel system, giving a readout time of 70 sec for 77 MB of data; the readout noise is $10e^-$, which is negligible since the sky contributes $\sim 3000e^-$/pixel. All the raw data is archived to Exabyte tape.

The typical exposure time is 300 sec, and about 60 images are taken per clear night; observations started in July 1992, and by July 1994 over 20,000 images had been taken.

### 2.1. Photometry

A special-purpose code is used to measure the brightness of all stars in the images; since we observe at standard sky positions, one high-quality frame for each image is used to define a 'template' catalogue of stars, and then subsequent reductions fit the brightness of the stars at known positions; this provides a major time saving as well as more accurate results. Many quality flags are stored for each data point, such as the $\chi^2$ of the fit to a stellar point spread function, and the fraction of the star affected by bad pixels, cosmic rays etc.

The photometric data is then rearranged into a time-series for each star, and an automated analysis is used to search both for variable stars and microlensing candidates. For the microlensing search, the timeseries is convolved with a set of filters of different durations. If a filter shows a peak at modest significance, a full 5-parameter fit to microlensing is made, and additional statistics are calculated. (The 5 parameters are the 2 shape parameters $A_{max}, \hat{t}$; the time of peak, and the un-amplified flux in our 2 colours). Stars satisfying modest cuts are then output separately, and subjected to more rigorous selection cuts to produce a final list of candidate events[6] .

## 3. LMC RESULTS

At present we have processed most of the first year's LMC data, comprising some 5500 frames distributed over 22 fields; this sample contains a total of 8 million stars observed for 1 year. This data has been searched for variable stars and microlensing candidates; some 40,000 variables have been found, most newly discovered. The majority of these fall into four well-known classes: there are approximately 20000 very red semi-regular or irregular variables, 7000 RR Lyraes, 2000 Cepheids and 1200 eclipsing binaries. Examples of such objects may be found in ref. [7]. We exclude the reddest 0.5% of our stars from the microlensing search as these stars can mimic long-timescale lensing events given limited time coverage, though they do not remain constant before or after the peak.

The microlensing search has yielded one striking candidate [8] with $A_{max} \approx 7.2$; [2] and two lower amplitude candidates (Figure 2). The first event has now been rediscovered from independent data in a field overlap, and shows good consistency between the two fields. A spectrum of this star has been measured [9]; it is a normal giant star, with a radial velocity consistent with that of the LMC. It is therefore very hard to explain this event by intrinsic stellar variability, given both the normality of the spectrum and the basic fact that the star's radius would have to expand by a factor $> 2.5$ with $\lesssim 5\%$ temperature change.

Clearly, the events 2 and 3 are much lower signal-to-noise than event 1, due both to the lower amplifications (2.0 and 1.52) and the fact that event 2 occurs in a fainter star. However, they are relatively well separated from the 'background' of random noise; and event 2 is also located in a field overlap, and again the two datasets agree. Thus it is unlikely that these events are due to noise, but intrinsic stellar variability remains a possibility. We have also obtained a spectrum of the star involved in event 3; due to telescope problems it is of low quality but shows no obvious peculiarities.

If these events are due to microlensing, rough estimates of the lens masses can be made from the durations. Although the observed $\hat{t}$ values of 34, 18 and 28 days are measured fairly accurately, $\hat{t}$ depends on the mass, distance, and transverse velocity of the lens, all of which are unknown. Given a model for the halo, statistical mass estimates can be made[4], which are approximately 0.12, 0.03 and $0.08 M_{\odot}$ for these 3 events; but we stress that the uncertainties are large, with '$2\sigma$' limits a factor of 10 either way. The mass estimate also depends on the population to which the lens belongs (e.g. dark halo or disk), though for the LMC line of sight this dependence is modest[10]. Thus the timescales are consistent with the lenses being either brown dwarfs or low mass stars; though the abundance of low mass stars is constrained by deep star counts[11]. Note that since the amplification distribution is predicted *a priori*, the amplification provides no mass information without independent knowledge of the impact parameter $b$.

[2] Due to a small improvement in the fitting routine, this is slightly higher than the value of 6.86 quoted in [8]

## 4. GALACTIC BULGE RESULTS

By a fortunate coincidence, the dense concentration of stars around the Galactic center , the 'Galactic bulge', is about 12 hours in right ascension from the LMC, hence it is well placed for observing when the LMC is too low in the sky. The bulge provides an interesting target for microlensing searches, for several reasons [12,13]:

1) The line of sight to the Bulge lies near the plane of the galactic disk and thus (independent of any dark matter) events must be seen from lensing by the known population of low-luminosity disk stars, hence providing a check of the experiment.
2) Observations of bulge microlensing may also provide a test of the controversial 'disk dark matter'[14], and of the poorly-known low-mass end of the stellar mass function.
3) Microlensing events by halo dark matter may also be seen towards the bulge, and could potentially be separated from disk lensing using the timescales; though due to the shorter distance of $\approx 8\,\mathrm{kpc}$, the expected optical depth is only $\sim 1.3 \times 10^{-7}$,

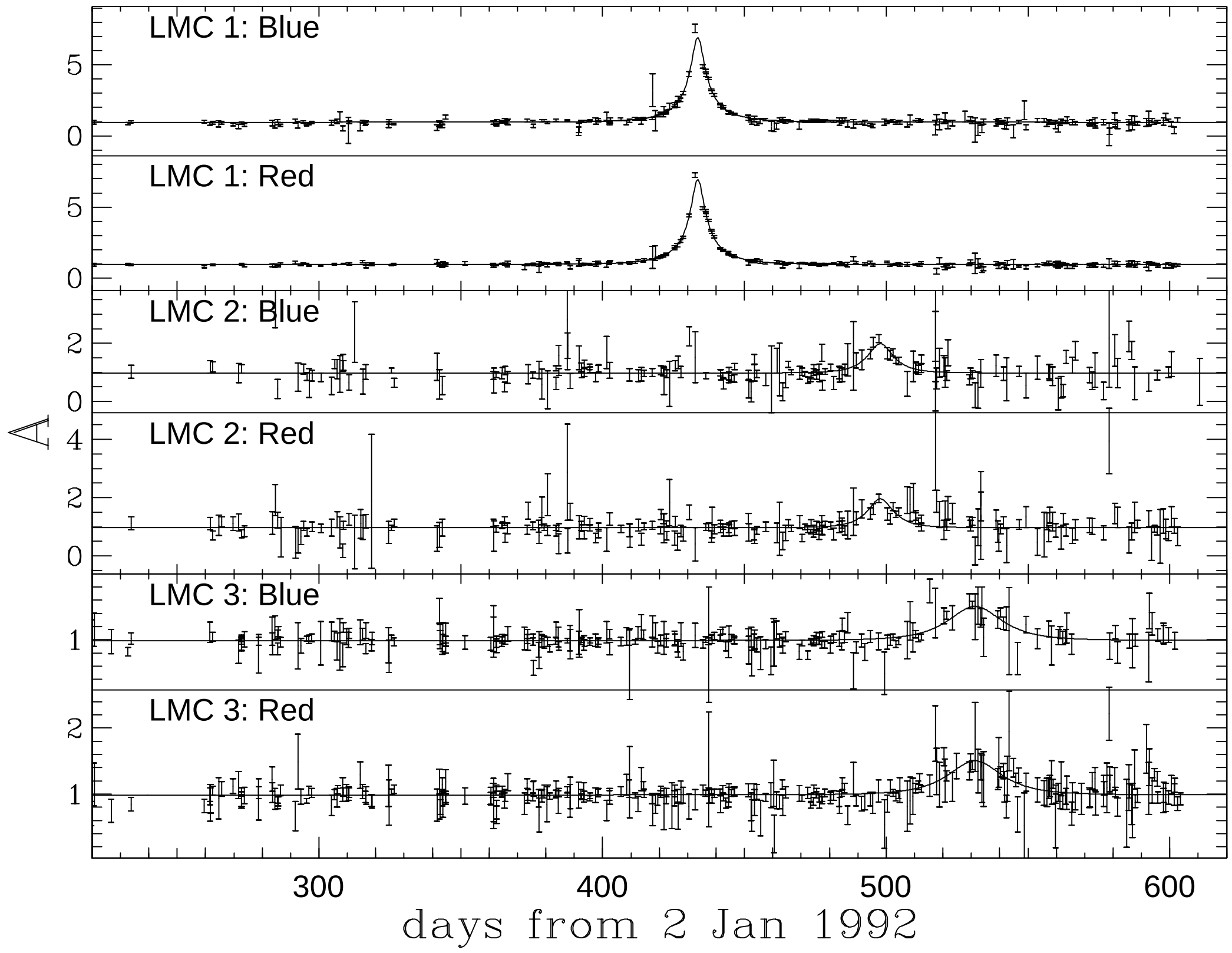

Figure 2. Light curves for the 3 candidate events in the LMC. Flux is in linear units, relative to the observed median flux. For each star the lower panel shows the red passband, the upper panel the blue. The smooth curves are the microlensing fit to both colours simultaneously.

and depends on the uncertain core radius of the dark halo.

At the time of this meeting, we had processed only one field's data from the Bulge, which unexpectedly showed four candidate microlensing events[6] (Figure 3), including one spectacular event with a peak amplification factor of $A_{max} > 17$. These four events are located in well-populated regions of the colour-magnitude diagram, and the amplifications correspond to dimensionless impact parameters of $u_{min} = 0.058, 0.36, 0.50, 0.66$, which is clearly consistent with the expected uniform distribution in $0 < u_{min} < 0.83$ (the upper limit set by the cut $A_{max} > 1.5$). The timescales of the events are in the range 20 – 48 days, which is approximately in the range expected from either low-mass disk stars or brown dwarfs.

Using Monte-Carlo models to set an upper limit $\mathcal{E} < 0.4$ on our detection efficiency, the predicted number of lensing events in this data due to disk stars is $\lesssim 1$, hence there is a significant excess of events[6]. This is consistent with results from the OGLE collaboration [15] .

This excess is unlikely to result from lensing by halo objects, as the optical depth is much too large, but there are two plausible explanations: 1) There may be a substantial contribution from lensing of bulge stars by other bulge stars; standard estimates of this optical depth [16] are too low to account for our observations, but if the bulge is elongated along the line-of-sight towards us, the optical depth could be substantially in-creased.

2) This optical depth could be accounted for by a 'maximal disk' which contains $\gtrsim 90\%$ of the Galaxy's mass interior to the Sun. Conventional estimates based on star-count data [14] and measurements of vertical motions[17] suggest a disk mass below half of this value, but it is possible that the extrapolation from local measurements is unreliable or that much of the disk mass is in brown dwarfs.

It is hard to discriminate these two possibilities by event timescales: since bulge lenses produce shorter events for a given mass due to the geometry, disk brown dwarfs and low-mass bulge stars give similar timescales. However, the variation of optical depth with galactic latitude and longitude provides a clear test: the optical depth to disk lensing depends on latitude only, while that for bulge lensing depends on galactic longitude also. This test will be applied in the next year.

### 4.1. Latest Status of Bulge Data

Between the meeting and the preparation of this manuscript, we have reduced and analysed another 10 bulge fields, containing over 5 million stars observed for 190 days. This has yielded approximately 23 additional events, proving that our first field is not anomalous; the distribution of $u_{min}$ is shown in Figure 4. One of these fields is in common with the 9 Baade's Window fields of the OGLE experiment[15], and we have independently rediscovered both their events (OGLE#1 and #7) from the 1993 season.

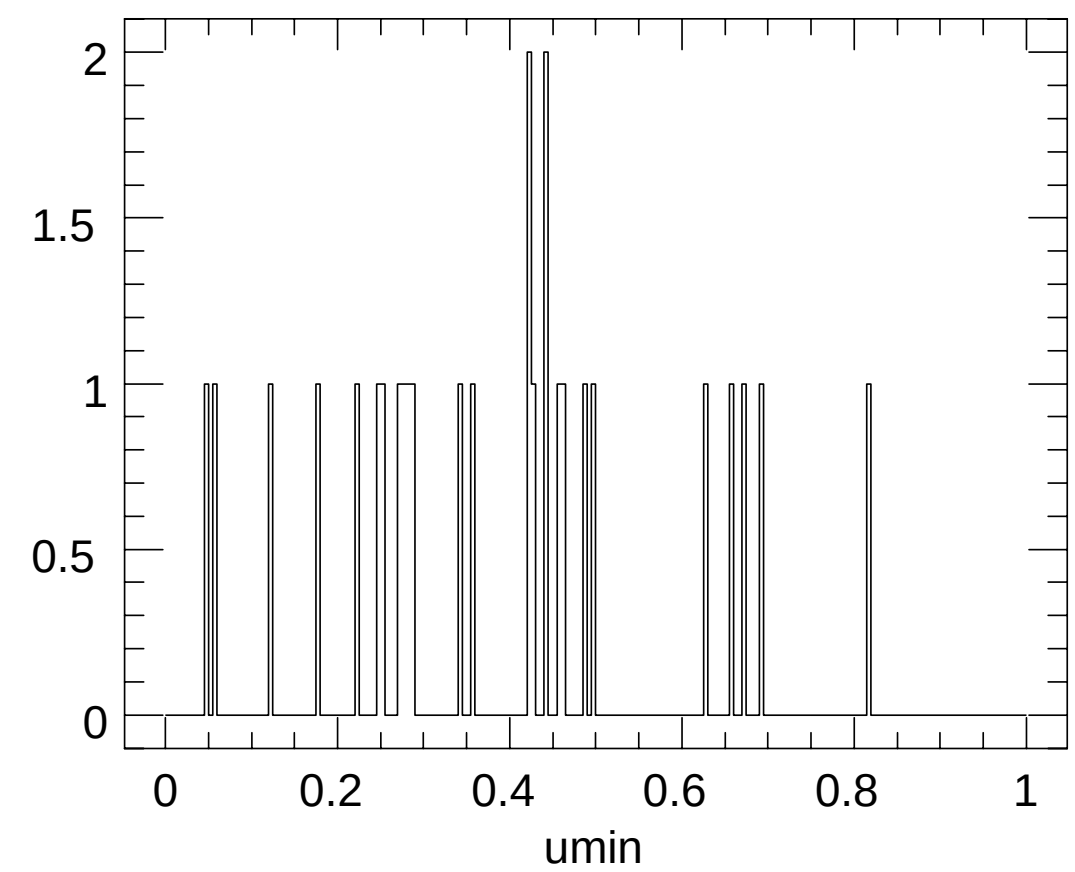

Figure 4. Histogram of dimensionless impact parameters $u_{min}$ derived from eq. (2) for 27 bulge lensing candidates.

## 5. DISCUSSION

It appears highly probable that most of the observed events are genuinely due to gravitational microlensing. The possibility of an observational artefact was never a serious concern due to the obvious brightening with a stellar PSF seen in many separate CCD frames; but our first LMC event[8] has been seen independently in two overlapping MACHO fields and the EROS data[18], firmly eliminating any possibility of observational error for this event.

For the lower-amplification events, the data is much less clear-cut; but if we accept that the high-amplification events are genuine microlensing, then lower amplification events must also occur. This suggests that at least the majority of our low-amplification candidates are also microlensing.

It is of course clear that the detected event rate towards the LMC is lower than naive theoretical predictions. For a standard dark halo of Machos of unique mass $M_L$ and an experimental efficiency $\mathcal{E}$, the number of events in our LMC dataset should be $\sim 40\mathcal{E}(M_L/0.1M_\odot)^{-0.5}$. However, although our efficiency estimates are still in progress it is already clear that $\mathcal{E} < 0.4$, and is likely to be below this when stellar crowding is accounted for. Given uncertainties in the halo models of probably a factor 2, it is premature to exclude a baryon-dominated halo.

Conversely, if our 3 events in the LMC are all microlensing, the rate is roughly consistent with the 2 EROS events[18], and appears significantly higher than expected from known stellar populations in the disk and spheroid[11]. It has been suggested that lensing by foreground LMC stars could provide most of the observed events[19], but this appears improbable[20]. Dark matter located in a thick disk[10] or spheroid[21] are possible al-ternative explanations. However, if some of the low-amplitude LMC events were not microlensing, the significance of the excess would be reduced; so at present it is not possible to make conclusive statements about the presence or absence of Machos in the Galaxy.

## 6. PROSPECTS

The prospects for the near future are very interesting: our observations will continue at least until 1996 to give much improved statistics; this will also check that our candidates do not repeat. A full efficiency calculation involving addition of fake stars to raw data frames is nearing completion, and will enable quantitative statements about the optical depth to the LMC. Spectroscopic observations of more of our candidate lensed stars will be obtained shortly.

There are also some possibilities for obtaining extra information about the lensing objects. If the lenses are low mass stars (as is probable for at least some of the bulge events) they may be detectable in high-quality spectra of the source stars, or in HST images a few years from now; though if they are brown dwarfs, there is little hope of direct observation.

We are currently testing a prototype real–time event detection system, with promising results. If events can be detected in real–time, we can obtain better photometry and spectra to check the lensing hypothesis; we have been allocated 1 hour per night on the CTIO 0.9-metre for follow-up photometry. Furthermore, small deviations from the point source approximation may occur due to e.g. the finite source size, or the motion of the Earth[10], and these can provide some additional constraints on the lens masses. Ideally one would like to measure event time-delays from a satellite in Solar orbit[22] to get the maximum information.

Although it is too early to make definite statements about the lensing objects, it is fairly clear that gravitational microlensing has been observed, and that we are not swamped by variable stars. Thus the MACHO project should soon be able to provide strong constraints on the abundance of compact objects in the disk and halo of our Galaxy.

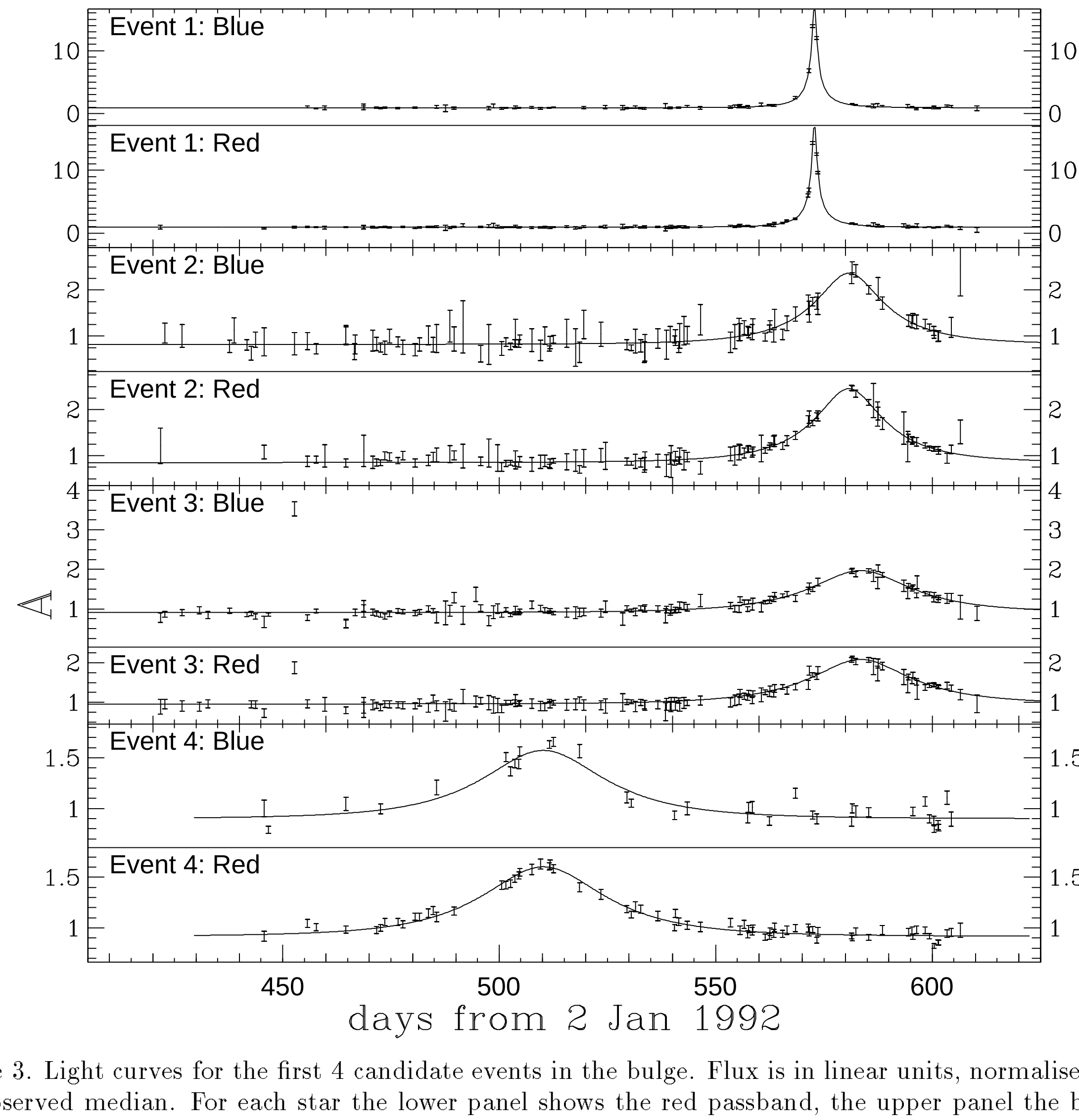

Figure 3. Light curves for the first 4 candidate events in the bulge. Flux is in linear units, normalised to the observed median. For each star the lower panel shows the red passband, the upper panel the blue. The smooth curves are the microlensing fits, simultaneous in both colours.